\documentclass[aps,eqsecnum,preprint,floats,epsf,epsfig,nofootinbib,showpacs]{revtex4}
\usepackage{graphicx}
\usepackage{amsmath}
\def\be{\begin{eqnarray}}
\def\en{\end{eqnarray}}
\def\non{\nonumber}

\def\bi{\bibitem}
\begin{document}

\title{\Large \bf Meson distribution amplitudes in holographic models
 }

\author{ \bf  Chien-Wen Hwang\footnote{
t2732@nknucc.nknu.edu.tw}}

\affiliation{\centerline{Department of Physics, National Kaohsiung Normal University,} \\
\centerline{Kaohsiung, Taiwan 824, Republic of China}
 }


\begin{abstract}
We study the wave functions of light and heavy mesons in both
hard-wall (HW) and soft-wall (SW) holographic models which use
AdS/CFT correspondence. In the case of massless constituents, the
asymptotic behaviors of the electromagnetic form factor, the
distribution amplitudes, and the decay constants for the two models are
the same, if the relation between the dilaton scale parameter and the
size of meson is an inverse proportion.
On the other hand, by introducing a quark mass dependence in the
wave function, the differences of the distribution amplitudes
between the two models are obvious. In addition, for the SW model, the
dependences of the decay constants of meson on the dilaton scale
parameter $\kappa$ differ; especially $f_{Qq}\sim
\kappa^3/m_Q^2$ is consistent with the prediction of the heavy quark
effective theory if $\kappa\sim m_Q^{1/2}$. Thus the parameters of
the two models are fit by the decay constants of the distinct
mesons; the distribution amplitudes and the $\xi$-moments are
calculated and compared.
\end{abstract}


\maketitle %
\section{Introduction}
The internal structure of the hadron has continually been a very important
and interesting subject in the research field of high-energy physics. From
the perspective of the experiment, in terms of the measurements
of decay rate or cross section, one can realize the internal
structure; on the other hand, from the viewpoint of theory, the internal
structure is displayed from the distribution amplitude or the form
factor. No matter whether decay or collision happens in the hadron, the
strong interaction plays an important role. Until now, the most
successful theory for describing the strong interaction is
quantum chromodynamics (QCD). However, it is still difficult
to carry out accurate calculations based on QCD; the main key lies in the
strong coupling constant, especially in the low-energy region,
close to unity. Such large value makes the perturbative
calculation an arduous process.

There were several nonperturbative approaches inspired by QCD which
could obtain some properties of the hadron and recently, based on the
correspondence of string theory in anti-de Sitter (AdS) space and
conformal field theory (CFT) in physical space-time
\cite{Ma,Gu,Witten,Ma2}, a semiclassical approximation to QCD,
light-front holography (LFH), was successfully developed for
describing the phenomenology of hadronic properties \cite
{Brodsky1,Brodsky2,Brodsky3,BT,Brodsky5,Brodsky6,Brodsky7,Brodsky8}.
LFH provides mapping of the string modes in AdS fifth dimension
$z$ to the light-front wave function in the impact variable $\zeta$,
which measures the separation of the constituents inside a hadron.
The mapping is proceeded by matching certain matrix elements
presented within the string theory in AdS space and the light-front
theory in Minkowski space. This approach, known as bottom-up, allows
to built models that have been successful in various QCD
applications, such as hadronic scattering processes
\cite{PS0,JP,LMKS}, hadronic spectrum \cite{KK,FBF,vega0,vega1},
hadronic couplings and chiral symmetry breaking
\cite{RP,EKSS,CFGJN}, and quark potentials \cite{FBF1,AZ,Jugeau}.

It is well known that the main feature of QCD as the fundamental
theory of the strong interaction lies in its nonperturbative behavior.
The picture of nonperturbative QCD leads to quark confinement and chiral
symmetry breaking, which are believed to be the two essential mechanisms
for forming hadron formation from QCD. Undoubtedly, the complicated and
nontrivial vacuum is not only the most important object to be understood
for the nonperturbative aspect of QCD, but also a starting point for the
construction of hadronic wave functions. Furthermore, in the light-
front (LF) coordinates, the nontrivial vacuum is connected to the particles
with zero-longitudinal momentum (the so-called zero-mode particles) and
the infrared divergences from the small longitudinal momentum \cite{Zhang}.
In fact, the LF infrared divergences are to be expected as the sources of
chiral symmetry breaking \cite{Zhang2}. However, the understanding and
application of the true QCD vacuum are still very limited in the LF
framework. Thus, here we study the relevant subjects in the LF framework
with a trivial vacuum. An alternate approach which can reveal the
phenomena of chiral symmetry breaking was proposed by the authors of
Ref. \cite{EKSS}, who chose the field content of the five-dimensional
theory to holographically reproduce the dynamics of chiral symmetry
breaking in QCD. While this holographic model depends on only three free
parameters, it agrees surprisingly well with the seven observables
of the light meson.

For compatibility with QCD, the conformal invariance must be
broken and the confinement in the infrared region must be
introduced. There are two types of the modified AdS geometry which can
achieve this results, as in the literature. One is the
``hard-wall" (HW) model \cite{Brodsky2,PS0,EKSS,RP,Brodsky6,GMTY,KLD,GR}
where partons are free inside the hadron and a hard cutoff is applied as
the boundary. Unlike the standard bag model \cite{bag}, these boundary
conditions are imposed on the impact variable $\zeta$, not on the bag
of the radius. The other is the ``soft-wall" (SW) model \cite{KK,AZ,FBF,GR1,
GK1,GK2,CKP,BSS,KL,CFGJN,vega0,vega1,GKK,vega2,GMNP,SWXY,AC,GHS,Afonin}
which has no sharp boundaries and employs a background dilaton field in
the AdS space as a smooth cutoff. 
A problem with the HW model is that the dependence of hadron masses
on the higher orbital angular momenta is linear, which is different from
the quadratic behavior or the so-called ``Regge trajectory". Conversely,
the SW model was initiated for solving the problem of the
hadronic mass spectrum. Each of these two models has certain
advantages which are explained the above references.

In this work, we study the wave functions of light and heavy mesons and
make the comparisons of the asymptotic behaviors of the electromagnetic
(EM) form factor, the distribution amplitudes, and the $\xi$-moments
in both the HW and SW holographic models. The case of the massless
constituents is first considered, and a generalization regarding the behavior
of the massive quarks \cite{BTmassive} follows. There are two points worth
mentioning: First, for the HW model, we use the parameter $l$ instead of
$1/\Lambda_{\textrm{QCD}}$ to represent the size of the meson in both cases of the
massless and massive constituents. Second, for the SW model, although both
dilaton field profiles $\textrm{exp}(\pm \kappa^2 z^2)$ have been used in the
literature \cite{Brodsky+,KK} to reproduce the behavior of Regge
trajectories for higher spin states, and the authors of Ref. \cite{vega11}
related these solutions by a canonical transformation, the main motivation
for using the dilaton field profiles $\textrm{exp}(+ \kappa^2 z^2)$ is its
seemingly better confinement properties \cite{AZ}. However, as shown in
Refs. \cite{KK,KK2}, on one hand, the dilaton exponential $\textrm{exp}
(+ \kappa^2 z^2)$ leads to the existence of a spurious massless scalar
mode in the model; on the other hand, the desired confinement properties
can be realized in the model with the dilaton exponential $\textrm{exp}
(- \kappa^2 z^2)$. Thus, here we apply the latter one to
both the hadronic field and the bulk-to-boundary propagator.

The remained of this paper is organized as follows. Section II reviews the
extraction of wave functions by the holographic mapping of the
light-front formulism to AdS string mode. In Sec. III, we
compare both models in regard to the massless and
massive constituents. In Sec. IV, we fit the dilaton scale parameter
and the radius of bag with the decay constants of the distinct
mesons and estimate the distribution amplitudes. The first four
$\xi$-moments in both models are compared with those of the other
theoretical calculations. Conclusions are presented in Sec. V. Other
useful derivations for the discussion are given in the appendices.
\section{Formulism}
\subsection{AdS wave equations}
The action for a spin-$J$ field $\Phi_J$ in AdS$_5$ space-time in
presence of a dilaton background field $\varphi(z)$ is given by \cite{Vega2010}
 \be
S_\Phi=\frac{(-1)^J}{2}\int d^4 x d
z\sqrt{g}e^{-\varphi(z)}(\partial_N\Phi_J\partial^N\Phi_J-\mu^2_J(z)\Phi^2_J),
\label{action}
 \en
where $\mu^2_J(z)=\mu^2_J+g_J\varphi(z)$ is a ``dressed"
five-dimensional mass because of the interaction of $\varphi(z)$
with $\Phi_J$. The mass $\mu_J$ is related to the conformal
dimension $\triangle$ by $(\mu_J R)^2=(\triangle-J) (\triangle-4+J)$
and the coupling $g_J$ will be fixed later. In addition,
$g=|$det$g_{MN}|$, $g_{MN}$ is the metric tensor with $M,N=0,1,2,
3,4$, and the AdS metric is defined as
 \be
ds^2=g_{MN}dx^Mdx^N=\frac{R^2}{z^2}(\eta_{\mu\nu} dx^\mu dx^\nu-dz^2),
 \en
with $R$ the horizon radius and
$\eta_{\mu\nu}=$diag$(1,-1,-1,-1)$. $\Phi_J$ is a rank tensor
field $\Phi(x^M)_{M_1 \cdots M_J}$ which is totally symmetrical in all
of its indices. Factoring out the plane wave along the
Poincar$\acute{e}$ coordinates $x^\mu$, we get:
 \be
 \Phi_P(x,z)_J=e^{-iP\cdot x}\Phi(z)_J,\label{phi}
 \en
with four-momentum $P_\mu$ and invariant hadronic mass
$P^\mu P_\mu=\mathcal{M}^2$, and taking the variation of
(\ref{action}), the AdS wave equation for the spin-$J$ field is:
 \be
 \left[-\frac{z^{3-2J}}{e^{-\varphi(z)}}\partial_z\left(\frac{e^{-\varphi(z)}}{z^{3-2J}}
 \partial_z\right)+\left(\frac{\mu_J(z) R}{z}\right)^2\right] \Phi(z)_J=\mathcal{M}^2
 \Phi(z)_J.\label{eqPhi}
 \en
We may assume that the hadron field $\Phi_P(x,z)_J$ minimally couples to a
massless vector field $A^M(x,z)$ with the action \cite{PS,HYS}:
 \be
 S_A=\int d^4xdz\sqrt{g}e^{-\varphi(z)}[-\frac{1}{4}F_{MN}F^{MN}+i
 g_5A^M\Phi_{P'}^*(x,z)_J
 \overset{\text{\tiny $\leftrightarrow$}}{\partial}_M\Phi_{P}(x,z)_J],\label{LA}
 \en
where $F^{MN}=\partial^M A^N-\partial^N A^M$ and $g_5$ is a five-dimensional effective
coupling constant. In the AdS space, the propagation of an EM probe can
polarize along Minkowski coordinates as:
 \be
 A_\mu(x,z)=\epsilon_\mu e^{-iq\cdot x}V(Q^2,z),\qquad A_z=0,\label{A}
 \en
where $Q^2=-q^2>0$ and $V(Q^2,z)$ is the bulk-to-boundary propagator
with $V(0,z)=V(Q^2,0)=1$, since we are normalizing the bulk
solutions to the total charge operator, and a boundary limit of the
external current $A_\mu(x,0)=\epsilon_\mu e^{-iq\cdot x}$. Taking
the variation of the first term in Eq. (\ref{LA}), the AdS wave
equation for the external current is:
 \be
 \left[-\frac{z}{e^{-\varphi(z)}}\partial_z\left(\frac{e^{-\varphi(z)}}{z}
 \partial_z\right)+z^2Q^2\right] V(Q^2,z)=0. \label{eqV}
 \en
\subsection{Light-front framework}

The Lorentz-invariant Hamiltonian equation for a relativistic bound-state
hadron is given by
 \be
P_\mu P^\mu |\psi(P)\rangle=\mathcal{M}^2  |\psi(P)\rangle,
 \en
where $P_\mu$ is the four-momentum in four-dimensional Minkowski
space. In the LF QCD, the hadron four-momentum $P_\mu$ can be
expressed as: $P=(P^+,P^-,\textbf{P}_\bot)$ with $P^\pm=P^0\pm P^3$
and $\textbf{P}_\bot=(P^1,P^2)$. The hadronic state, which is an
eigenstate of $P^+$, $\textbf{P}_\bot$ and the total longitudinal
spin $J_z$, is normalized as
 \begin{equation}
\langle \psi (P^+,\textbf{P}_\bot, J_z)|\psi(P'^+,\textbf{P}'_\bot,
J'_z)\rangle=2(2\pi)^3 P^+\delta_{J_z,J'_z}
\delta(P^+-P'^+)\delta^2(\textbf{P}_\bot-\textbf{P}'_\bot).\label{norm}
 \end{equation}
In terms of the LF relative momentum variables $x_i$ and
$\textbf{k}_{\bot i}$, the on-shell partonic momentum $p_i$ can be
expressed as
 \be
p^+_i=x_i P^+,\quad \textbf{p}_{i\bot}=x_i
\textbf{P}_\bot+\textbf{k}_{\bot i},
 \en
and the hadronic state is expanded as
 \be
|\psi(P^+, \textbf{P}_\bot, J_z)\rangle &=&
\sum_{n,\lambda_i}\prod^n_{i=1}\int\frac{dx_id^2\textbf{k}_\bot}{2(2\pi)^3\sqrt{x_i}}
(16\pi^3)\delta(1-\sum^n_{j=1}x_j)\delta^2(\sum^n_{j=1}\textbf{k}_{\bot j}) \nonumber \\
&\times &\psi_n(x_i,\textbf{k}_{\bot
i},\lambda_i)|n;p_i,\lambda_i\rangle, \label{expandH}
 \en
where the delta functions are required from the momentum
conservation, $|n\rangle$ is the multiparticle Fock eigenstates with
$n$ the number of patrons in a given Fock state, and $\lambda_i$ is
the projection of the constituent's spin along $z$ direction. The
LF wave function $\psi_n(x_i,\textbf{k}_{\bot
i},\lambda_i)$ complies with the angular momentum sum rules \cite{BHMS}
$J_z=\sum^n_{i=1} \lambda_i+\sum^{n-1}_{i=1}L_{i,z}$ and is
normalized as
 \be
 \sum_{n,\lambda_i}\prod^n_{i=1}\int\frac{dx_id^2\textbf{k}_{\bot i}}{2(2\pi)^3\sqrt{x_i}}
 (16\pi^3)\delta(1-\sum^n_{j=1}x_j)\delta^2(\sum^n_{j=1}\textbf{k}_{\bot
 j})|\psi_n(x_i,\textbf{k}_{\bot i},\lambda_i)|^2=1,
 \en
which is derived from (\ref{norm}).

In the LF framework, a lot of physical quantities relate to the LF
wave function. Here we show some of them which are of concern. First,
the decay constant of a pseudoscalar meson $f_M$ is given by:
 \be
 \langle0|J^+_W|M(P^+,\textbf{P}_\bot)\rangle=iP^+f_M,\label{fm}
 \en
where $J^+_W = \bar {q}_1\gamma^+(1-\gamma^5)q_2$ is the flavor
changing weak current which is evaluated at fixed light-cone time $x^+=0$.
Only the valence quarks contribute to the decay; thus, we expand the initial state in
Eq. (\ref{fm}) into the Fock component and find
 \be
 f_M=2\sqrt{2 N_c}\int \frac{d x d^2 \textbf{k}_\bot}{2(2\pi)^3}\psi_{\bar
 {q} q}(x,\textbf{k}_\bot).\label{fmexpand}
 \en
Equation (\ref{fmexpand}) enables the straight evaluation of the decay constant in terms
of the LF wave function $\psi_{\bar {q} q}(x,\textbf{k}_\bot)$.

Next, the distribution amplitude $\phi(x,Q)$, which is the amplitude for finding
constituents with longitudinal momentum fraction $x$ in the meson and collinear up to the scale
$Q$, is defined as \cite{Brodsky80}:
 \be
 \phi(x,Q)\equiv \int^{Q^2}
 \frac{d^2\textbf{k}_\bot}{2(2\pi)^3}\psi_M(x,\textbf{k}_\bot).\label{distribution}
 \en
The meson state $|M\rangle$, which is expanded into Fock states, will approximate
$|\bar{q}q\rangle$ when $Q^2$ is large. Therefore, we can also calculate
$\phi(x)\equiv\phi(x,Q\to\infty)$ by the wave function $\psi_{\bar {q} q}
(x,\textbf{k}_\bot)$. In addition, an approach to parameterize distribution
amplitude is to calculate the so-called $\xi$-moments:
 \be
 \langle \xi^N \rangle=\frac{\int^1_{-1}d\xi \xi^N \phi(\xi)}{\int^1_{-1}d\xi \phi(\xi)},
 \label{ximoment}
 \en
where $\xi=1-2 x$.

Finally, the EM form factor of a meson is defined as:
 \be
 \langle M(P')|J^+(0)|M(P)\rangle=(P+P')^+F(Q^2),\label{em}
 \en
where $J^+(y)=\sum_q e_q \bar {q}(y) \gamma^+ q(y)$, $P'=P+q$, and
$F(0)=1$. Using the expanded state Eq. (\ref{expandH}) and the
normalization condition Eq. (\ref{norm}), the EM form factor of a
meson can be expressed as \cite{dyw1,dyw2,soper}:
 \be
 F(Q^2)&\equiv& \int dx \rho(x, \textbf{q}_\bot),\non \\
 &=&\int\frac{dx d^2\textbf{k}_\bot}{2(2\pi)^3}\sum_i e_i \psi^*_{\bar{q}q}(x,
 \bar{\textbf{k}}_\bot)\psi_{\bar{q}q}(x,\textbf{k}_\bot), \label{Fq2}
 \en
where $\rho(x, \textbf{q}_\bot)$ is an effective single-particle density and
$\bar{\textbf{k}}_\bot=\textbf{k}_\bot+(1-x_i) \textbf{q}_\bot$. Equation (\ref{Fq2}) reveals
that the current matrix element in Eq.(\ref{em}) can be represented as overlaps of
the LF wave functions.

\subsection{Light-front holographic mapping}
For the holographic mapping of the light-front wave function to AdS
string mode, it is convenient to define the transverse center of
momentum of a hadron $\textbf{R}_\bot$ as:
 \be
 \textbf{R}_\bot=\frac{1}{P^+}\int dy^- d^2\textbf{y}_\bot T^{++}
 \textbf{y}_\bot,
 \en
where $T^{\mu\nu}$ is the energy momentum tensor. Then the partonic
transverse position $\textbf{r}_{\bot i}$ and the internal
coordinates $\textbf{b}_{\bot i}$, which conjugate to the relative
momentum variable $\textbf{k}_{\bot i}$, have the following
relations:
 \be
 x_i \textbf{r}_{\bot i}=x_i \textbf{R}_\bot + \textbf{b}_{\bot i},
 \quad \sum^n_{i=1} \textbf{b}_{\bot i}=0,
 \quad \textbf{R}_\bot=\sum^n_{i=1} x_i \textbf{r}_{\bot i}.
 \en
The light-front wave function $\psi_n(x_i,\textbf{k}_{\bot i})$ can
be expressed by the internal coordinates $\textbf{b}_{\bot i}$ as:
 \be
 \psi_n(x_i,\textbf{k}_{\bot
 i})=(4\pi)^{(n-1)/2}\prod^{n-1}_{i=1}\int d^2\textbf{b}_{\bot i}
 \textrm{exp}(i\sum^{n-1}_{i=1}\textbf{b}_{\bot i}\cdot \textbf{k}_{\bot
 i})\tilde{\psi}_n(x_i,\textbf{b}_{\bot i}),\label{fourier}
 \en
where $\tilde{\psi}_n(x_i,\textbf{b}_{\bot i})$ is the light-front
wave function in the coordinate space and is normalized as:
 \be
 \sum_{n}\prod^n_{i=1}\int dx_id^2\textbf{b}_{\bot i}
 |\tilde{\psi}_n(x_i,\textbf{b}_{\bot i})|^2=1.
 \en
Then, we can substitute Eq. (\ref{fourier}) into Eq. (\ref{Fq2}) and integrate over
$\textbf{k}_\bot$ phase space. The EM form factor of a meson $F(Q^2)$ 
can be obtained as:
 \be
 F(Q^2)=\int dx d^2\textbf{b}_\bot\textrm{exp}(i\textbf{q}_\bot\cdot (1-x)
 \textbf{b}_\bot)|\tilde{\psi}_{\bar {q} q}(x,\textbf{b}_\bot)|^2.
 \en
We can also express $F(Q^2)$ in terms of an effective single-particle transverse
distribution $\tilde{\rho}(x,\textbf{c}_\bot)$ \cite{soper}:
 \be
 F(Q^2)=\int dx d^2 \textbf{c}_\bot \textrm{exp}(-i\textbf{c}_\bot \cdot
 \textbf{q}_\bot)\tilde{\rho}(x,\textbf{c}_\bot),\label{Fc}
 \en
where $\textbf{c}_\bot=(1-x)\textbf{b}_\bot$ is the $x$-weighted transverse position
coordinate of the spectator quark, and
 \be
 \tilde{\rho}(x, \textbf{c}_\bot)&=&\int\frac{d^2 \textbf{q}_\bot}{(2\pi)^2}\textrm{exp}
 (i\textbf{c}_\bot \cdot \textbf{q}_\bot)\rho(x,\textbf{q}_\bot)\non \\
 &=&\int d^2 \textbf{b}_\bot\delta^2((1-x)\textbf{b}_\bot-\textbf{c}_\bot)|\tilde{\psi}
 _{\bar{q}q}(x,\textbf{b}_\bot)|^2.\label{rhoLF}
 \en
If we integrate Eq. (\ref{Fc}) over angles, the form factor can be obtained as \cite{BT}:
 \be
 F(Q^2)=2\pi \int dx\frac{1-x}{x}\int \zeta d\zeta J_0\left(\zeta q \sqrt{\frac{1-x}{x}}\right)
 \tilde{\rho}(x, \zeta),\label{Fzeta}
 \en
where $\zeta = \sqrt{x(1-x)}|\textbf{b}_\bot|$ and $J_n$ is the Bessel function of the
first kind.

On the other hand, if we consider a pseudoscalar meson coupled to an external EM field in
AdS space, the second term of Eq. (\ref{LA}) can be related to a hadron matrix element as
\cite{PS,BT}:
 \be
 &&\int d^4xdz\sqrt{g}e^{-\varphi(z)}A^M\Phi_{P'}^*(x,z)_{J'}
 \overset{\text{\tiny
 $\leftrightarrow$}}{\partial}_M\Phi_{P}(x,z)_J\non \\
 &&\qquad \sim(2\pi)^4\delta^4(P'-P-q)
 \epsilon_\mu\langle M(P',J')|J^\mu|M(P,J)\rangle. \label{correspond}
 \en
Substituting Eqs. (\ref{phi}) and (\ref{A}) to the left hand side of Eq. (\ref{correspond}) and
extracting a delta function from momentum conservation at the vertex, the form factor can be
obtained as:
 \be
 F(Q^2)=R^3\int\frac{dz}{z^3}e^{\varphi(z)}\Phi(z)V(Q^2,z)\Phi(z), \label{Fz}
 \en
and be treated as the overlap in the fifth dimension coordinate $z$
of the normalizable modes which correspond to the
incoming and outgoing mesons, $\Phi_P$ and $\Phi_{P'}$, with the non-normalizable mode,
$V(Q^2,z)$, which deal to the external source. Comparing Eq. (\ref{Fz})
with Eq. (\ref{Fzeta}) and identifying the holographic variable $z$
with the transverse LF variable $\zeta$, the holographic mapping of
the light-front wave function to AdS string mode can be accomplished. Next,
two kinds of dilaton background fields were considered and their
respective wave functions extracted.

\subsubsection{Hard Model}
The simplest dilaton background field is none; that is,
$\varphi(z)=0$. Thus, the AdS wave equation, Eq. (\ref{eqV}), for
the external current is reduced as:
 \be
 [z^2\partial^2_z-z \partial_z-z^2Q^2]V_\textrm{H}(Q^2,z)=0,
 \en
and its solution with the boundary condition is:
 \be
 V_\textrm{H}(Q^2,z)=z Q K_1(z Q),\label{Vsolution}
 \en
where $K_\alpha(x)$ is the modified Bessel functions of the second kind. In addition,
an integral is done \cite{BT} as:
 \be
 \int^1_0 dx J_0\left(\zeta Q \sqrt{\frac{1-x}{x}}\right)=\zeta Q K_1(\zeta Q).
 \label{J0K1}
 \en
Substituting Eqs. (\ref{Vsolution}) and (\ref{J0K1}) to Eqs. (\ref{Fz}) and (\ref{Fzeta}),
respectively, and comparing the latter two equations, the relation between the LF wave
function and the AdS string mode is obtained as:
 \be
 |\tilde{\psi}_{\bar{q}q}(x,\zeta)|^2=\frac{R^3}{2\pi}x(1-x)\frac{|\Phi_\textrm{H}(\zeta)|^2}{\zeta^4},
 \label{hardrelation}
 \en
where $z=\zeta$  and Eq. (\ref{rhoLF}) are applied. On the other hand, for the meson field,
Eq. (\ref{eqPhi}) is reduced as:
 \be
 \left[-\partial^2_z+\frac{3-2J}{z} \partial_z+\left(\frac{\mu_J R}{z}\right)^2\right]
 \Phi_\textrm{H}(z)=\mathcal{M}^2\Phi_\textrm{H}(z).\label{mesonH}
 \en
By substituting $\tilde{\Phi}_\textrm{H}(\zeta)=(\zeta/R)^{-3/2+J}\Phi_\textrm{H}(\zeta)$, an
effective Schrodinger equation is obtained as:
 \be
 \left[-\frac{d^2}{d\zeta^2}+\frac{4
 L^2-1}{4\zeta^2}\right]\tilde{\Phi}_\textrm{H}(\zeta)=\mathcal{M}^2\tilde{\Phi}_\textrm{H}(\zeta),\label{phieq}
 \en
where the conformal dimension $\Delta=L+2$ and $(\mu_J
R)^2=L^2-(2-J)^2$ are used. However, for the known confinement
inside a hadron, a baglike model \cite{bag} is considered where
partons are free inside the meson and are forbidden outside the
meson. An additional hard-well potential is needed: $U(\zeta)=0$ if
$\zeta\leq l$ and $U(\zeta)=\infty$ if
$\zeta\geq l$, where $l$ is the size of hadron. Then, the solution to Eq.
(\ref{phieq}) is $\tilde{\Phi}_\textrm{H}(\zeta)=C \zeta^{1/2} J_L(\zeta \mathcal{M})$
with a boundary condition $\tilde{\Phi}_\textrm{H}(\zeta=l)=0$ and a normalization condition
$\int d\zeta |\tilde{\Phi}(\zeta)|^2=1$. As for the meson field,
the solution to Eq. (\ref{mesonH}) is:
 \be
 \Phi_\textrm{H}(z)=\frac{z^{2-J}}{R^{3/2-J}}\frac{\sqrt{2}}{l J_{L+1}(\mathcal{M}l)}
 J_L(z \mathcal{M}).\label{PhiH}
 \en
From Eq. (\ref{hardrelation}),
the LF wave function in the HW model in the limit of massless
constituents is:
 \be
 \tilde{\psi}_{L,n}(x,\textbf{b}_\bot)=\frac{1}{\sqrt{\pi}l J_{L+1}(\beta_{L,n})}\sqrt{x(1-x)}
 J_L(\sqrt{x (1-x)}|\textbf{b}_\bot| \mathcal{M}_{L,n}) \theta\left(\textbf{b}_\bot^2 \leq
 \frac{l^2}{x (1-x)}\right),\non \\ \label{hardb}
 \en
where $\beta_{L,n}$ is the root of the Bessel function, $\mathcal{M}_{L,n}=\beta_{L,n}/l$, and $\theta$
is the step function. 
\subsubsection{Soft Model}
Another background field which introduced an infrared soft cutoff is a dilaton $\varphi(z)=
\kappa^2 z^2$. This time Eq. (\ref{eqV}) is reduced as:
 \be
 [z^2\partial^2_z-(1+2 \kappa^2 z^2) z \partial_z-z^2Q^2]V_\textrm{S}(Q^2,z)=0,
 \en
and the solution with the boundary condition is:
 \be
 V_\textrm{S}(Q^2,z)=\Gamma\left(1+\frac{Q^2}{4 \kappa^2}\right)U\left(\frac{Q^2}{4 \kappa^2},0,\kappa^2z^2
 \right),\label{softV}
 \en
where $\Gamma(x)$ is the Gamma function and $U(a,b,c)$ is the confluent hypergeometric function. In
the large $Q^2$ limit, that is, $Q^2\gg4 \kappa^2$, the solution is reduced as \cite{BT}:
 \be
 V_\textrm{S}(Q^2,z)\rightarrow z Q K_1(z Q)=V_\textrm{H}(Q^2,z).\label{LQ}
 \en
Thus, in the large $Q$ limit, the relation between the LF wave function and the AdS string mode is
obtained as:
 \be
 |\tilde{\psi}_{\bar{q}q}(x,\zeta)|^2=\frac{R^3}{2\pi}x(1-x)e^{-\kappa^2z^2}\frac{|\Phi_\textrm{S}(\zeta)|^2}{\zeta^4}.
 \label{softrelation}
 \en
For this background field, Eq. (\ref{eqPhi}) is reduced as:
 \be
 \left[-\partial^2_z+\frac{3-2J+2 \kappa^2 z^2}{z} \partial_z+\left(\frac{\mu_J(z) R}{z}\right)^2\right]
 \Phi(z)=\mathcal{M}^2\Phi(z),
 \en
and an effective Schrodinger equation can be obtained as:
 \be
 \left[-\frac{d^2}{d\zeta^2}+\frac{4
 L^2-1}{4\zeta^2}+\kappa^4 z^2+\kappa^2[g_J R^2-2(J-1)]\right]\tilde{\Phi}_\textrm{S}(\zeta)=\mathcal{M}^2
 \tilde{\Phi}_\textrm{S}(\zeta),\label{phieqsoft}
 \en
where $\tilde{\Phi}_\textrm{S}(\zeta)=e^{-\kappa^2\zeta^2/2}(\zeta/R)^{-3/2+J}\Phi_\textrm{S}(\zeta)$.
The normalized solution to Eq. (\ref{phieqsoft}) is:
 \be
 \tilde{\Phi}_\textrm{S}(\zeta)=\kappa^{L+1}\sqrt{\frac{2 n!}{(n+L)!}}\zeta^{L+1/2}e^{-\kappa^2\zeta^2/2}L^L_n(\kappa^2\zeta^2),
 \en
with
 \be
 \mathcal{M}^2=4 \kappa^2\left(n+\frac{L-J+2+g_J R^2/2}{2}\right),
 \en
and the meson field is:
 \be
 \Phi_\textrm{S}(z)=\frac{z^{2-J+L}}{R^{3/2-J}}\kappa^{L+1}\sqrt{\frac{2 n!}{(n+L)!}}L^L_n(\kappa^2z^2).\label{softPhi}
 \en
In order to obtain a massless pion and a linear Regge trajectories: $\mathcal{M}^2\sim J$ at large $J$, the value
$g_J R^2=4(J-1)$ is fixed. Thus, from Eq. (\ref{softrelation}), the LF wave function in the soft-wall model in the
limits of massless constituents is:
 \be
 \tilde{\psi}_L(x,\textbf{b}_\bot)=\frac{\kappa^{L+1}}{\sqrt{\pi}}\sqrt{\frac{n!}{(n+L)!}}[x (1-x)]^{(L+1)/2}|
 \textbf{b}_\bot|^L e^{-\kappa^2x (1-x)\textbf{b}_\bot^2/2}L^L_n(\kappa^2x (1-x)\textbf{b}^2_\bot),\label{softb}
 \en
with
 \be
 \mathcal{M}^2=4 \kappa^2\left(n+\frac{L+J}{2}\right).
 \en
\section{Two holographic models in comparison}
\subsection{massless constituents}
In this section, we make the comparisons between the HW and SW models for some hadron properties.
For the EM form factor of the charged pion, we take $n=J=L=0$ and substitute Eqs. (\ref{Vsolution}), (\ref{PhiH})
and (\ref{softV}), (\ref{softPhi}) into Eq. (\ref{Fz}):
 \be
 F^\textrm{H}_\pi(Q^2)&=&2 \int dz \frac{z}{l^2J^2_1(\beta_{0,1})}J_0^2\left(\frac{z \beta_{0,1}}{
 l}\right)z Q K_1(z Q), \non \\
 F^\textrm{S}_\pi(Q^2)&=&2 \int dz z\kappa^2 (L_0^0(\kappa^2 z^2))^2\Gamma\left(1+\frac{Q^2}{4 \kappa^2}\right)
 U\left(\frac{Q^2}{4 \kappa^2},0,\kappa^2z^2\right),\label{Fpian}
 \en
which correspond to the HW and SW models,
respectively. For the HW model, the analytical result is not easily obtained.
As for the SW model, the EM form factor is \cite{BT}:
 \be
 F^\textrm{S}_\pi(Q^2)=\frac{4 \kappa^2}{4\kappa^2+Q^2}.\label{softFpi}
 \en
However, if we check the asymptotic behavior in the large $Q$ limit, the results are
 \be
 F_\pi(Q^2\rightarrow\infty)\rightarrow -\frac{R^3}{Q^2}\frac{e^{\varphi(z)}\Phi(z)^2}{z^4}\Bigg|^{Q u}_0.
 \en
where $u=l(\infty)$ for the HW(SW) model. Thus:
 \be
 Q^2F^\textrm{H}_\pi(Q^2)|_{Q^2\rightarrow\infty}=4\left(\frac{1}{l J_1(\beta_{0,1})}\right)^2,\qquad
 Q^2F^\textrm{S}_\pi(Q^2)|_{Q^2\rightarrow\infty}=4\kappa^2.
 \en
They are derived in Appendix A and, of course, the latter is easily
checked according to Eq. (\ref{softFpi}). In other words, if we set a
parametric relation:
 \be
 \kappa=\frac{1}{l J_1(\beta_{0,1})},\label{kappaLambda}
 \en
the asymptotic behaviors of $Q^2F_\pi(Q^2)$ for both models are the
same.
In addition, the mean square radius of the meson $P$ is determined from the slope of $F_P(Q^2)$ at $Q^2 = 0$:
 \be
 \langle r^2_P\rangle=-6\frac{d F_P(Q^2)}{dQ^2}\bigg|_{Q^2=0}.
 \en
Thus, from Eq. (\ref{softFpi}), the mean square radius of the pion in the SW model is
 \be
 \langle r^2_\pi\rangle^\textrm{S}=\frac{3}{2\kappa^2}.\label{msrpi}
 \en
For the HW model, although the value of $\langle r^2_\pi\rangle^\textrm{H}$ diverges logarithmically, as mentioned
in Ref. \cite{BT}, this problem in defining mean square radius of the pion does not appear if one uses Neumann
boundary conditions, and $\langle r^2_\pi\rangle^\textrm{H}\sim l^2$. Thus, the inverse
relation between $\kappa$ and $l$ is still satisfied.

One may speculate whether or not there are other similarities between these two models.
Thus, we consider $\tilde{\Phi}_{\textrm{H,S}}(\zeta)$, which are the normalized solutions of the effective
Schrodinger equation and are derived from the meson field, for $L=0$. 
If the parametric relation (\ref{kappaLambda}) is used, we have:
 \be
 \tilde{\Phi}^{n=0}_{\textrm{H}}(\zeta)=\sum_{n=0}^\infty c_n \tilde{\Phi}^n_{\textrm{S}}(\zeta),\label{expand}
 \en
where $\sum^\infty_0 |c_n|^2=1$. It should be noted that $c_n$ is independent of the parameters
$\kappa$ and $l$, which is described in Appendix B. Qualitatively, the coefficient
$c_0$ tells us ``how much" $\tilde{\Phi}^0_{\textrm{S}}(\lambda)$ is contained in
$\tilde{\Phi}^0_{\textrm{H}}(\lambda)$. We easily find $c_0\simeq0.980$. In other words, the former is
dominated within the latter. 

It is convenient to study other hadron properties in the momentum space. The LF wave function of the HW
model for $n=L=0$ can be obtained from Eqs. (\ref{fourier}) and (\ref{hardb}) as:
 \be
 \psi_{\textrm{H}}(x,\textbf{k}_\bot)=\frac{4\pi\sqrt{x (1-x)}\beta_{0,1}l}{x(1-x)\beta^2_{0,1}
 -k^2l^2}J_0\left(\frac{k l}{\sqrt{x (1-x)}}\right),\label{wfH}
 \en
where $k=|\textbf{k}_\bot|$. The distribution amplitude and decay constant
can be obtained from Eqs. (\ref{distribution}) and (\ref{fmexpand}), 
respectively:
 \be
 \phi_{\textrm{H}}(x)=\frac{\sqrt{x(1-x)}}{2\pi l J_1(\beta_{0,1})},\qquad f_{\textrm{H}}=
 \frac{\sqrt{6}}{8 l J_1(\beta_{0,1})}, \label{masslessH}
 \en
and the $\xi$-moments can be obtained by Eq. (\ref{ximoment}):
 \be
 \langle \xi^N \rangle_{\textrm{H}}=\frac{1+(-1)^N}{2\sqrt{\pi}}\frac{\Gamma[\frac{N+1}{2}]}{\Gamma[\frac{N}{2}+2]}.
 \en
We see that the $\xi$-moments is zero when $N$ is odd. In other words, the distribution amplitude is symmetric
for $\xi=0$ or $x=\frac{1}{2}$. This is apparent because of $\phi_{\textrm{H}}(x)\propto \sqrt{x(1-x)}$.

\begin{figure}
 \includegraphics*[width=4in]{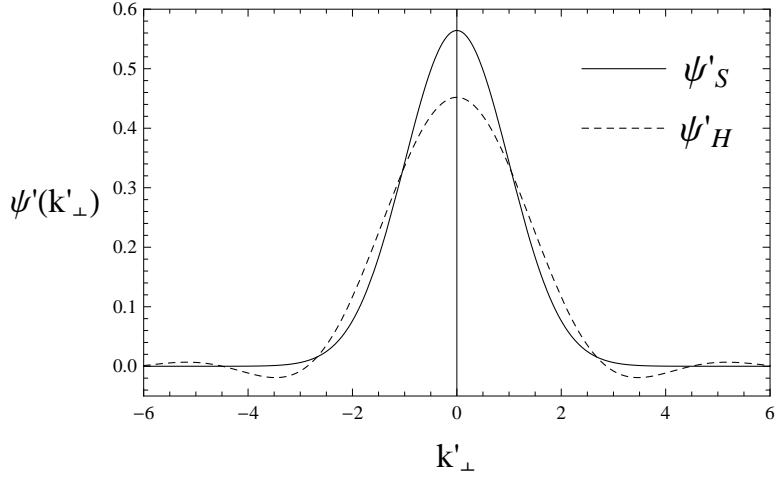}
 \caption{Profile of the LF wave functions for the massless constituents. The
 solid and dotted lines correspond to $\psi'_{\textrm{S}}(\textbf{k}'_\bot)$ and
 $\psi'_{\textrm{H}}(\textbf{k}'_\bot)$, respectively.}
  \label{fig:c00}
 \end{figure}

On the other hand, also for $L=0$, the LF wave function of the SW model in momentum space can be obtained from Eqs.
(\ref{fourier}) and (\ref{softb}) as:
 \be
 \psi^n_{\textrm{S}}(x,\textbf{k}_\bot)=\frac{(-1)^n 4\pi}{\kappa\sqrt{x(1-x)}}e^{-\frac{k^2}{2 x (1-x) \kappa^2}}
 L^0_n\left(\frac{k^2}{x (1-x) \kappa^2}\right),
 \label{wfS}
 \en
and the distribution amplitude and decay constant for $n=0$ 
can be obtained as:
 \be
 \phi_{\textrm{S}}(x)=\frac{\sqrt{x(1-x)}\kappa}{2\pi},\qquad f_{\textrm{S}}=
 \frac{\sqrt{6}\kappa}{8}.\label{masslessS}
 \en
Equations (\ref{masslessH}) and (\ref{masslessS}) are consistent with the
result of Ref. \cite{BT} where they used a derivation:
$\phi(x,\mathcal{Q}\rightarrow \infty)\rightarrow
\tilde{\psi}(x,\textbf{b}_ \bot\rightarrow 0)/\sqrt{4\pi}$ as
$\zeta\rightarrow 0$. The $\xi$-moments of this distribution
amplitude, because of the same dependence as that of the hard-wall
model: $\phi_{\textrm{S}}(x)\propto \sqrt{x(1-x)}$, is $\langle
\xi^N \rangle_{\textrm{S}}=\langle \xi^N \rangle_{\textrm{H}}$. In
fact, if we again set the parametric relation in Eq.
(\ref{kappaLambda}), then:
 \be
 \phi_{\textrm{S}}(x)=\phi_{\textrm{H}}(x),\qquad f_{\textrm{S}}=f_{\textrm{H}}.
 \label{HSequal}
 \en
As for the similarity between $\psi_{\textrm{H}}(x,\textbf{k}_\bot)$ and $\psi_{\textrm{S}}
(x,\textbf{k}_\bot)$, we can follow the similar argument for $\tilde{\Phi}_{\textrm{H,S}}(\zeta)$
in Appendix B:
introducing a new variable $\textbf{k}'_\bot=\textbf{k}_\bot/\sqrt{x (1-x)}\kappa$,
and rewriting the LF wave functions of the two models as
 \be
 \psi'_{\textrm{H}}(\textbf{k}'_\bot)=\frac{\beta_{0,1}}{\sqrt{\pi}(\beta^2_{0,1}J_1(\beta_{0,1})
 -k'^2/J_1(\beta_{0,1}))}J_0\left(\frac{k'}{J_1(\beta_{0,1})}\right),
 \en
and
 \be
 \psi'_{\textrm{S}}(\textbf{k}'_\bot)=\frac{1}{\sqrt{\pi}}e^{-\frac{k'^2}{2}},
 \en
where $k'=\mid\textbf{k}'_\bot\mid$ and both new functions satisfy
the normalization condition: $\int d^2\textbf{k}'_\bot
\mid\psi'(\textbf{k}'_\bot)\mid^2=1$. We may consider a complete set:
$\{\psi'^{\textrm{HO}}_{n}(\textbf{k}'_\bot)\}$ which is the
solution of the Schr\"{o}dinger equation for a two-dimension
harmonic oscillator:
 \be
 \psi'^{\textrm{HO}}_{n}(\textbf{k}'_\bot)=\frac{(-1)^n}{\sqrt{\pi}}e^{-\frac{k'^2}{2}}L^0_n(k'^2).
 \en
Here the function
$\psi'_{\textrm{S}}(\textbf{k}'_\bot)=\psi'^{\textrm{HO}}_0(\textbf{k}'_\bot)$, which is the
``ground state" for this two-dimension harmonic oscillator, and the function
$\psi'_{\textrm{H}}(\textbf{k}'_\bot)$ can be written as a linear
combination of $\{\psi'^{\textrm{HO}}_n(\textbf{k}'_\bot)\}$: $\psi'_{\textrm{H}}(\textbf{k}'_\bot)=\sum
c'_n\psi'^{\textrm{HO}}_n(\textbf{k}'_\bot)$.
We can easily check that $c'_0\simeq0.980$, $c'_n=(-1)^n c_n$, and $\sum c'^2_n=1$. The coefficients $c'_n$ also
satisfy the relation:
 \be
 \psi_{\textrm{H}}(\textbf{k}_\bot)=\sum_n c'_n\psi^n_{\textrm{S}}(\textbf{k}_\bot).\label{HSrelation}
 \en
This means that the SW wave function is dominated within the HW one
for $n=L=0$. We can realize this situation in terms of sketching $\psi'_{\textrm{S}}(\textbf{k}'_\bot)$
and $\psi'_{\textrm{H}}(\textbf{k}'_\bot)$ in Fig. \ref{fig:c00}. 
If the integration of $\textbf{k}_\bot$ for both sides of Eq. (\ref{HSrelation}) is taken, we obtain
 \be
 \frac{\sqrt{x (1-x)}}{2 \pi l J_1(\beta_{0,1})}=\sum_n c'_n \frac{\sqrt{x (1-x)}\kappa}{2\pi},
 \en
where
 \be
 \int \frac{d^2 \textbf{k}_\bot}{16\pi^3} \psi^n_{\textrm{S}}(\textbf{k}_\bot)= \frac{\sqrt{x (1-x)}\kappa}{2\pi},
 \en
is independent of $n$. Thus, using the parametric relation Eq. (\ref{kappaLambda}), we have another constraint of
$c'_n$:
 \be
 \sum^\infty_0 c'_n=1.\label{cnp}
 \en
The convergence of $\sum c'_n$ is slower than that of $\sum c'^2_n$ because the sign of $c'_n$ may be positive or negative.
\subsection{massive constituents}
A simple generalization of
the LF wave function for massive quarks follows from the assumption
that the momentum space LF wave function is a function of the
invariant off-energy shell quantity \cite{BTmassive}:
 \be
 \mathcal{M}^2-\varepsilon=\frac{\textbf{k}^2_\bot+m^2_1}{x}+\frac{\textbf{k}^2_\bot+m^2_2}{1-x}\equiv M_0^2.
 \en
In other words, one may make the following replacement:
 \be
 \frac{k^2}{x(1-x)}\rightarrow M^2_0=\frac{k^2}{x(1-x)}+m^2_{12},
 \qquad m^2_{12}=\frac{m^2_1}{x}+\frac{m^2_2}{1-x},
 \label{m12}
 \en
where 
$m_i$ is the current mass of quark. 
For the SW model, this replacement is equivalent to a change of the kinetic term in the effective
Schrodinger equation Eq. (\ref{phieqsoft}) \cite{Vega2010}:
 \be
 -\frac{d^2}{d\zeta^2}\rightarrow  -\frac{d^2}{d\zeta^2}+m^2_{12}.\label{add}
 \en
The alterant masses of the light and heavy mesons were obtained by combining Eqs.
(\ref{phieqsoft}) and (\ref{add}) and solving them:
 \be
 M^2_{n_J}=4\kappa^2\left(n+\frac{L+J}{2}\right)+N \int^1_0 dx~m^2_{12} e^{-\frac{m^2_{12}}{\kappa^2}},
 \en
where $N$ is the normalization constant fixed from the integral $N \int dx e^{-\frac{m^2_{12}}{\kappa^2}}=1$.

Here we generalize this replacement to the
HW and SW models. Thus, the LF wave functions of these two models for $n=0$ can be replaced as:
 \be
 \psi_{\textrm{H}}(x,\textbf{k}_\bot,m_i )&\sim&\frac{4\pi \beta_{0,1}l}{\sqrt{x (1-x)}
 \left(\beta^2_{0,1}- M_0^2 l^2\right)}J_0(M_0 l),\non \\
 \psi_{\textrm{S}}(x,\textbf{k}_\bot,m_i)&\sim&\frac{4\pi}{\kappa\sqrt{x(1-x)}}e^{-\frac{M_0^2}{2\kappa^2}}.
 \label{wfm2}
 \en
If the distribution amplitude for massive quarks is considered,
we find that they are just the integrations of the massless LF wave function with an infrared cutoff:
 \be
 \phi_\textrm{H,S}(x,m_i)\equiv\int^\infty_{-\infty} \frac{d^2\textbf{k}_\bot}{2(2\pi)^3}\psi_\textrm{H,S}
 (x, \textbf{k}_\bot,m_i)\sim\int^\infty_\mu \frac{k dk}{2(2\pi)^2}\psi_\textrm{H,S}(x,\textbf{k}_\bot),
 \en
where $\mu=\sqrt{x (1-x)} m_{12}$. For the SW model, we obtain
 \be
 \phi_\textrm{S}(x,m_i)\sim\frac{\sqrt{x (1-x)} \kappa}{2\pi}e^{\frac{-m^2_{12}}{2 \kappa^2}},
 \en
which is consistent with the result of \cite{Vega2010}. As for the HW model, we may use Eq. (\ref{HSrelation}) to
obtain:
 \be
 \phi_\textrm{H}(x,m_i)\sim\sum_n c'_n\int^\infty_\mu \frac{k dk}{2(2\pi)^2}\psi^n_\textrm{S}(x,\textbf{k}_\bot).
 \en
Thus, the distribution amplitude for the HW model is 
 \be
 \phi_\textrm{H}(x,m_i)&\sim&\frac{\sqrt{x (1-x)} \kappa}{2\pi}e^{\frac{-m^2_{12}}{2 \kappa^2}}\Bigg[c'_0+c'_1
 \left(1+\frac{m^2_{12}}{\kappa^2}\right)+c'_2\left(1+\frac{m^4_{12}}{2\kappa^4}\right)\non \\
 &+&c'_3\left(1+\frac{m^2_{12}}
 {\kappa^2}-\frac{m^4_{12}}{2\kappa^4}+\frac{m^6_{12}}{6\kappa^6}\right)+c'_4\left(1+\frac{m^4_{12}}
 {\kappa^4}-\frac{m^6_{12}}{3\kappa^6}+\frac{m^8_{12}}{24\kappa^8}\right)
 +\ldots\Bigg].\label{phiH-S}
 \en
Because of the constant term of $L^0_n(t)$, for all $n$'s, being equal to $1$, we substitute Eq. (\ref{cnp}) into
Eq. (\ref{phiH-S}) and obtain:
 \be
 \phi_\textrm{H}(x,m_i)&\sim&\phi_\textrm{S}(x,m_i)\Bigg[1+c'_1
 \frac{m^2_{12}}{\kappa^2}+c'_2\frac{m^4_{12}}{2\kappa^4}
 +c'_3\left(\frac{m^2_{12}}
 {\kappa^2}-\frac{m^4_{12}}{2\kappa^4}+\frac{m^6_{12}}{6\kappa^6}\right)\non \\
 &&\qquad\qquad+c'_4\left(\frac{m^4_{12}}
 {\kappa^4}-\frac{m^6_{12}}{3\kappa^6}+\frac{m^8_{12}}{24\kappa^8}\right)+\ldots\Bigg].\label{comparephi}
 \en
Thus, the difference between $\phi_\textrm{H}(x,m_i)$ and $\phi_\textrm{S}(x,m_i)$ is displayed as a function of
$m^2_{12}/\kappa^2$. 
If the ratio $m^2_{12}/\kappa^2\ll 1$, that is, the meson is composed of the light quarks, the predictive
distribution amplitudes from the HW and SW models are nearly the same. In contrast, if $m^2_{12}/\kappa^2\geq 1$,
the parameter relation $\kappa=1/l J_1(\beta_{0,1})$ may be unable to be satisfied and the distribution amplitudes will be
quite different. We will display these comparisons in the next section.

In addition, for the SW model, if we consider the heavy and light quarkonium states, the scalings of the leptonic
decay constants are:
 \be
 f_{QQ}\sim \frac{\kappa^{3/2}}{m_Q^{1/2}},\qquad
 f_{qq}\sim \frac{\sqrt{6}\kappa}{8}-\left(a_1+a_2 \ln\left[\frac{m_q}{\kappa}\right]\right)\frac{m_q^2}{\kappa},
 \en
which is described in Appendix C. The latter is consistent with the case of the massless quark in Eq.
(\ref{masslessS}): $f\sim \kappa$. On the other hand, if we take the heavy quark limit $(m_Q\rightarrow\infty)$,
the scaling of the leptonic decay constants of heavy-light mesons is:
 \be
 f_{Qq}\sim \frac{\kappa^3}{m_Q^2},
 \en
which is described in Appendix D. This result is in agreement with the heavy quark effective theory
(HQET):
 \be
 f_\textrm{HQET}\sim 1/m_Q^{1/2},\label{HQET}
 \en
if $\kappa\propto m_Q^{1/2}$.
These results reveal that the dilaton scale parameter $\kappa$ seems
to vary with the quark mass; this inference is in accordance with
the conclusion in Section III. D of Ref. \cite{Vega2010}, even
though they fixed $\kappa$ and varied an additional parameter
$\lambda_{qQ}$ with the distinct mesons.
\section{Numerical Results and Discussion}
\begin{figure}
 \includegraphics*[width=4in]{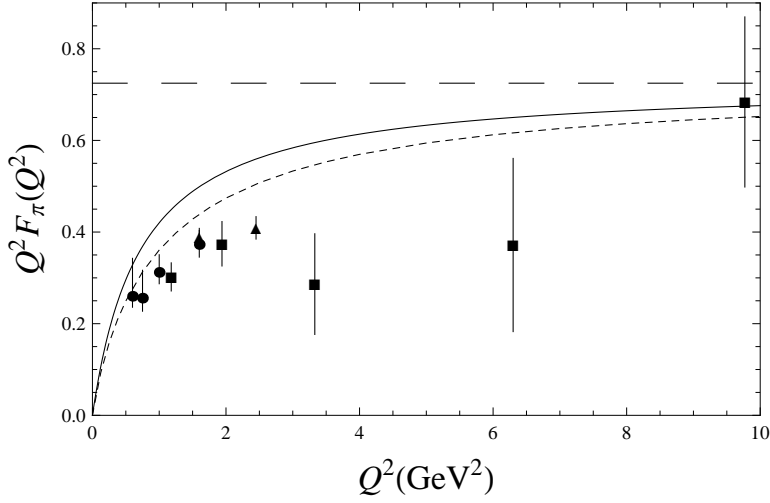}
 \caption{$Q^2F_\pi(Q^2)$ as a function of $Q^2$. The solid and dotted lines correspond to the SW and HW models,
 respectively. Data are taken from \cite{Jdata1} (triangles), \cite{Jdata2} (circles), and \cite{dataL} (boxes) for
 large $Q$ transfers. The long dashed line is the limiting behavior $4\kappa^2$.}
  \label{fig:Fpi}
 \end{figure}

For the pion, whose constituent quarks approximate masslessness, the numerical results of the EM form factor for the HW
and SW models are obtained by Eq. (\ref{Fpian}) and the parameters $l_\pi=0.892$ fm, $\kappa_\pi=0.426$ GeV,
which fit the decay constant of pion $f_{\textrm{H}}=f_{\textrm{S}}=f_\pi=130.41$ MeV \cite{data}, and shown in
Fig. \ref{fig:Fpi}.
We find that the difference between the two lines is large for the region $1 \textrm{GeV}^2 \leq Q^2 \leq 4
\textrm{GeV}^2$, while for the others, it is relatively small. The long dashed line in Fig. \ref{fig:Fpi} is the limiting behavior
$4\kappa^2$ for both models. In addition, the line for the SW model is a little different from that in Fig. 2
of \cite{BT} because for the latter, the parameter $\kappa=0.375$ GeV is obtained by fitting the
data for the form factor. However, for the parameter $\kappa_\pi=0.426$ GeV, the mean square radius of pion $\langle r^2_\pi
\rangle^\textrm{S}=0.32~\textrm{fm}^2$, which is small compared with the PDG value $\langle r^2_\pi
\rangle=0.45~\textrm{fm}^2$ \cite{data}.

\begin{figure}
 \includegraphics*[width=4in]{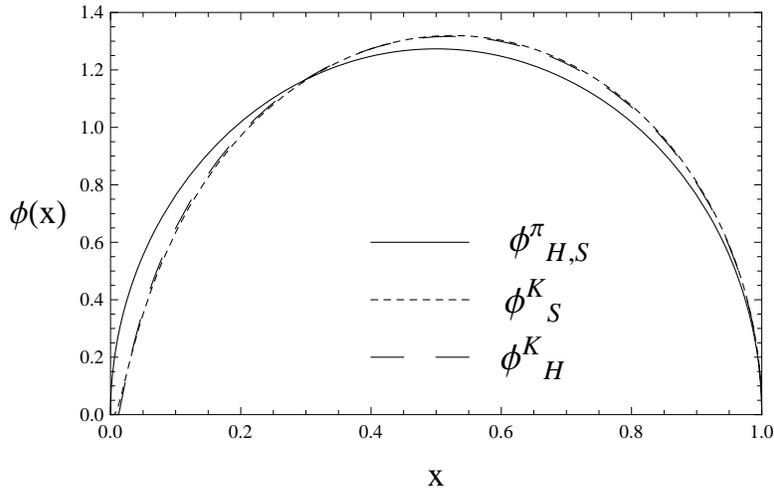}
 \caption{Distribution amplitudes of the light mesons. The solid line
 corresponds to the $\pi$ meson for the SW and HW models. The long dashed
 and dotted lines correspond to the $K$ meson for the SW and HW models,
 respectively. The latter two are nearly the same.}
  \label{fig:light}
 \end{figure}

\begin{figure}
 \includegraphics*[width=4in]{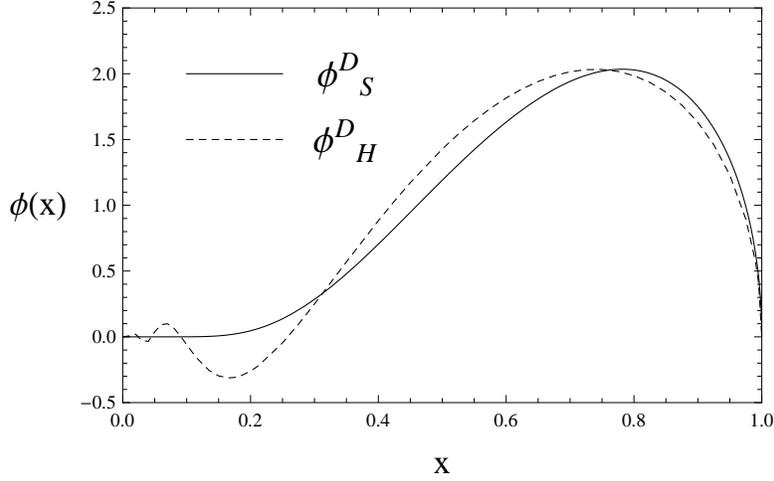}
 \caption{Distribution amplitudes of $D$ meson. The solid and dotted lines correspond to the SW and HW models,
 respectively.}
  \label{fig:D}
 \end{figure}

\begin{figure}
 \includegraphics*[width=4in]{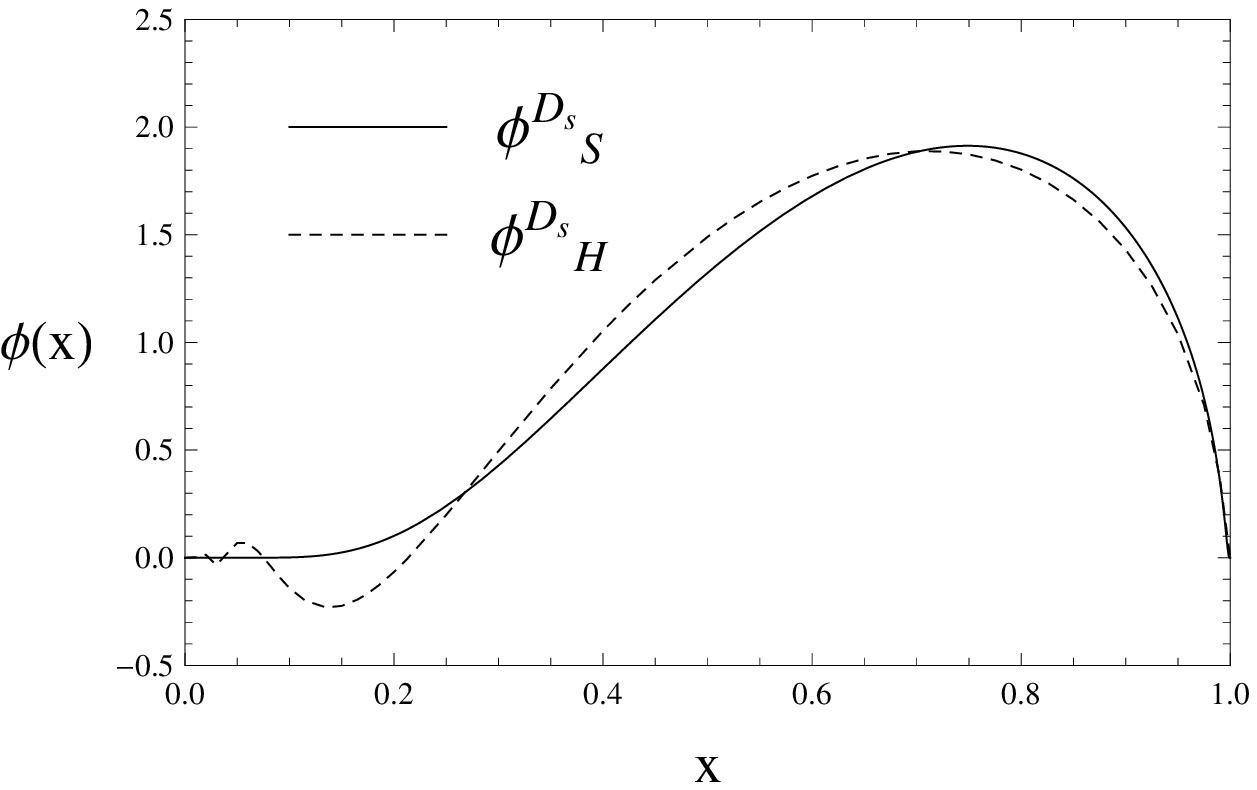}
 \caption{Distribution amplitudes of $D_s$ meson. The solid and dotted lines correspond to the SW and HW models,
 respectively.}
  \label{fig:Ds}
 \end{figure}

\begin{figure}
 \includegraphics*[width=4in]{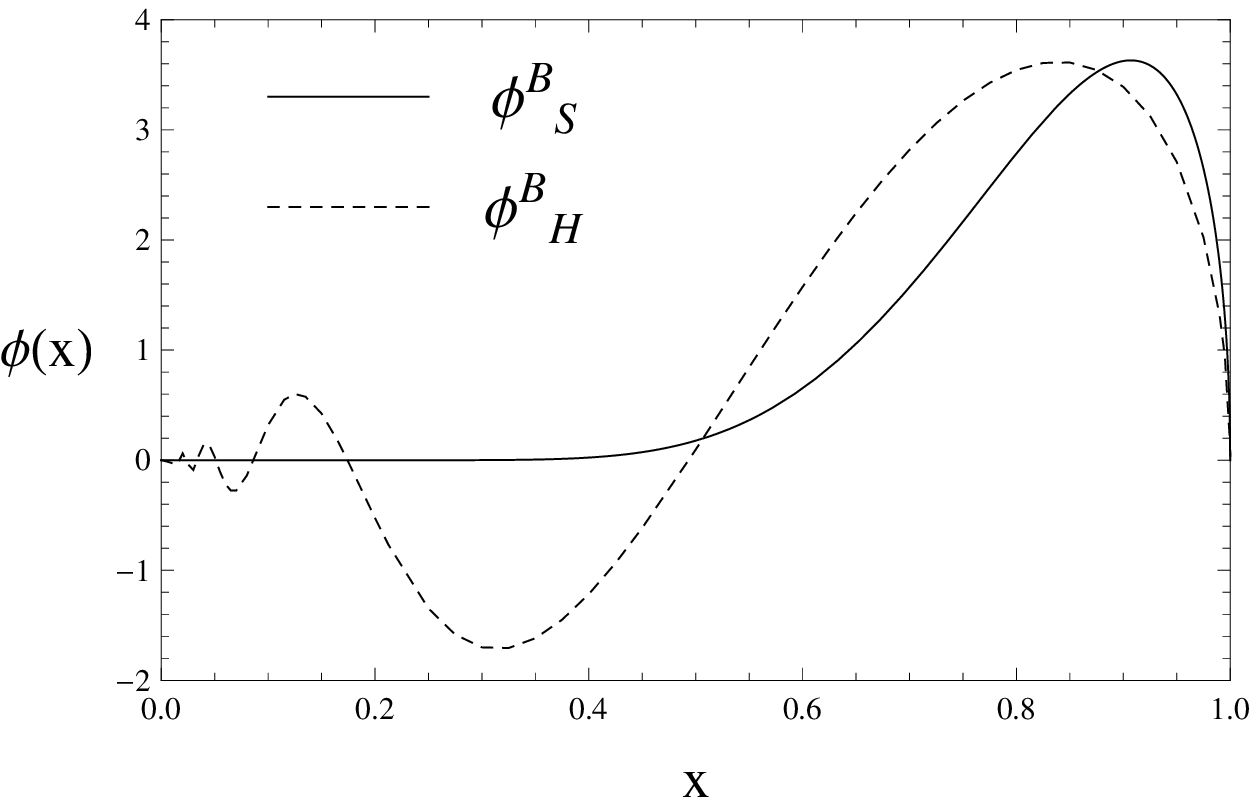}
 \caption{Distribution amplitudes of $B$ meson. The solid and dotted lines correspond to the SW and HW models,
 respectively.}
  \label{fig:B}
 \end{figure}

\begin{figure}
 \includegraphics*[width=4in]{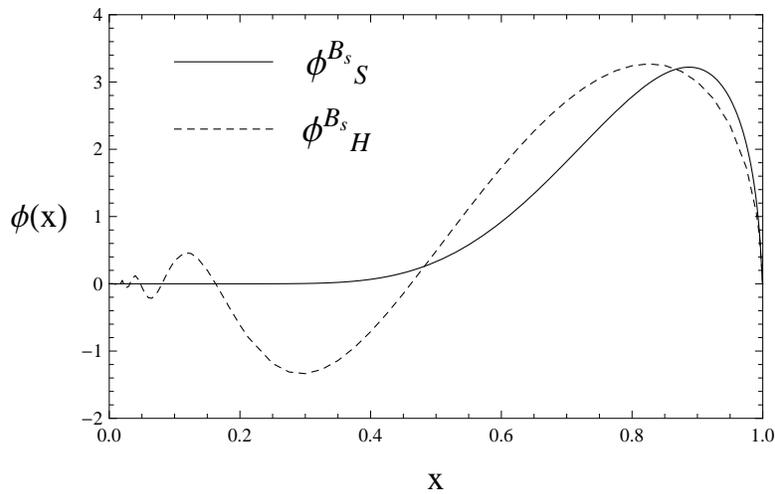}
 \caption{Distribution amplitudes of $B_s$ meson. The solid and dotted lines correspond to the SW and HW models,
 respectively.}
  \label{fig:Bs}
 \end{figure}

\begin{figure}
 \includegraphics*[width=4in]{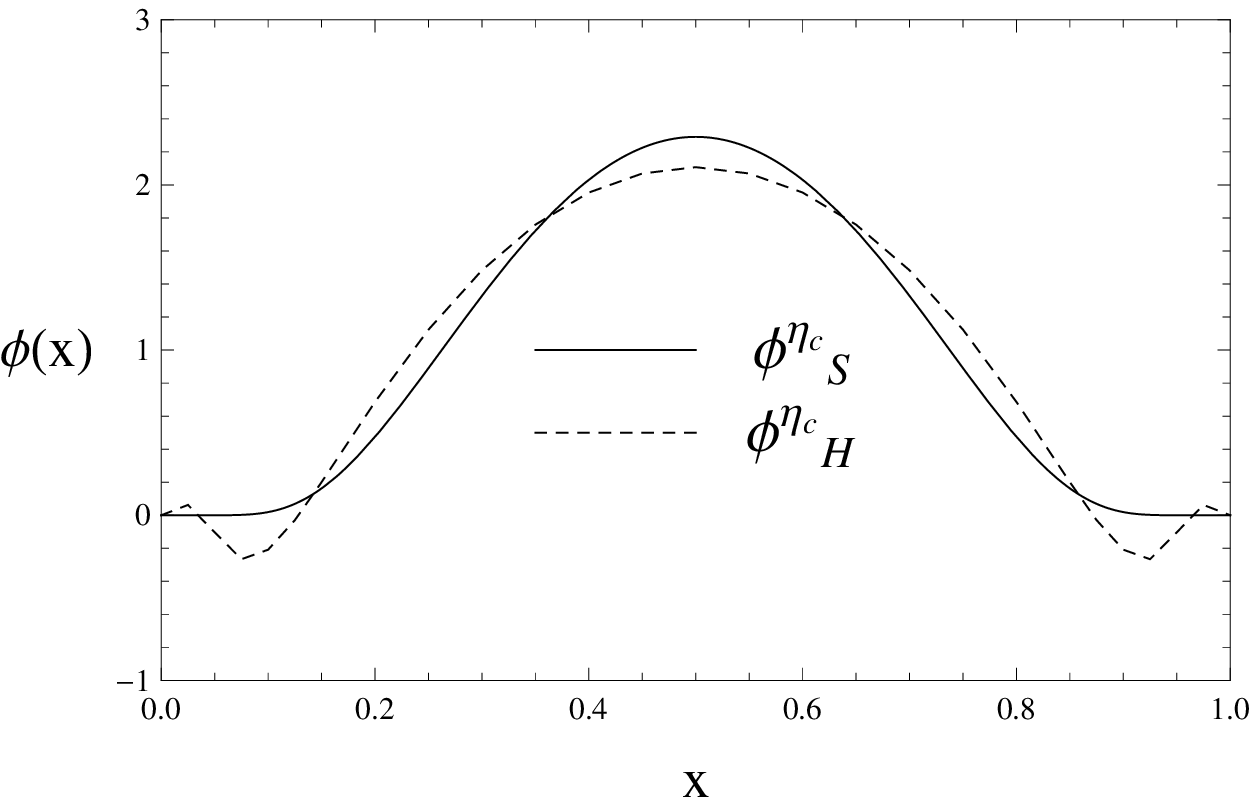}
 \caption{Distribution amplitudes of $\eta_c$ meson. The solid and dotted lines correspond to the SW and HW models,
 respectively.}
  \label{fig:etac}
 \end{figure}

\begin{figure}
 \includegraphics*[width=4in]{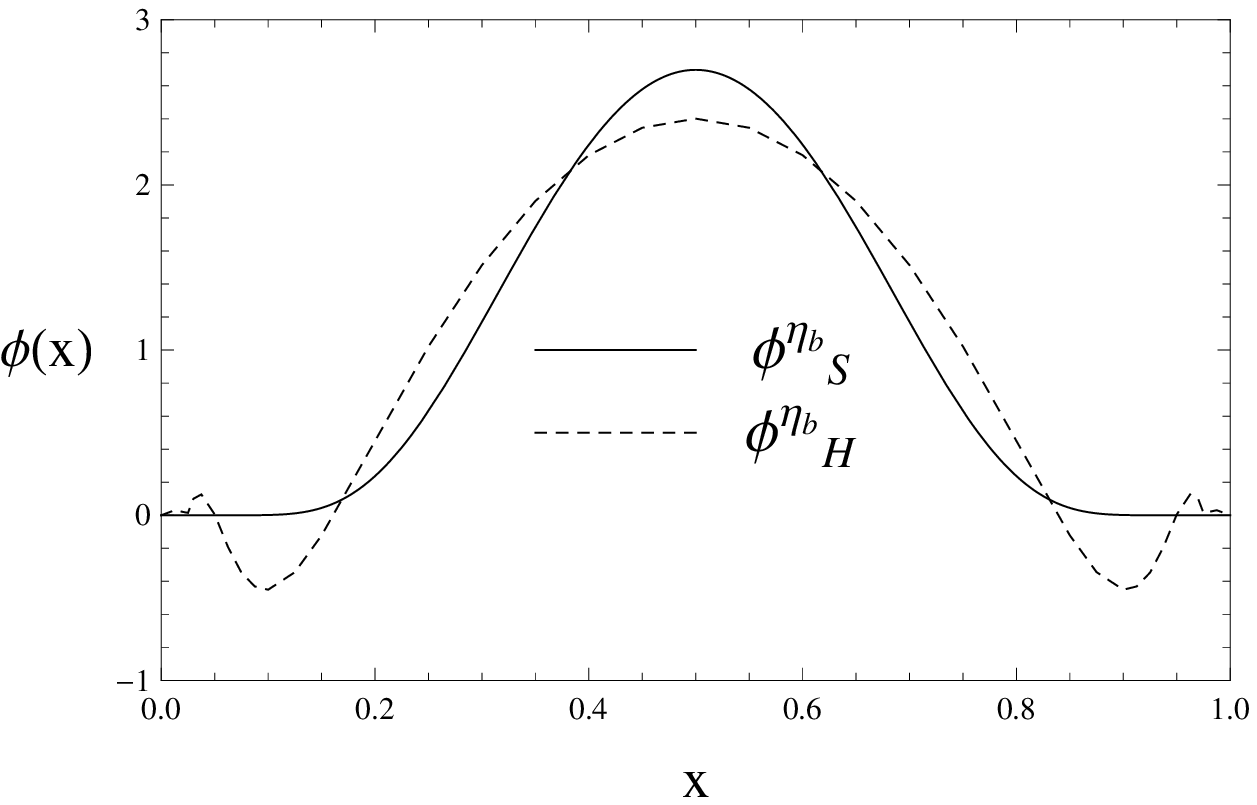}
 \caption{Distribution amplitudes of $\eta_b$ meson. The solid and dotted lines correspond to the SW and HW models,
 respectively.}
  \label{fig:etab}
 \end{figure}

\begin{figure}
 \includegraphics*[width=4in]{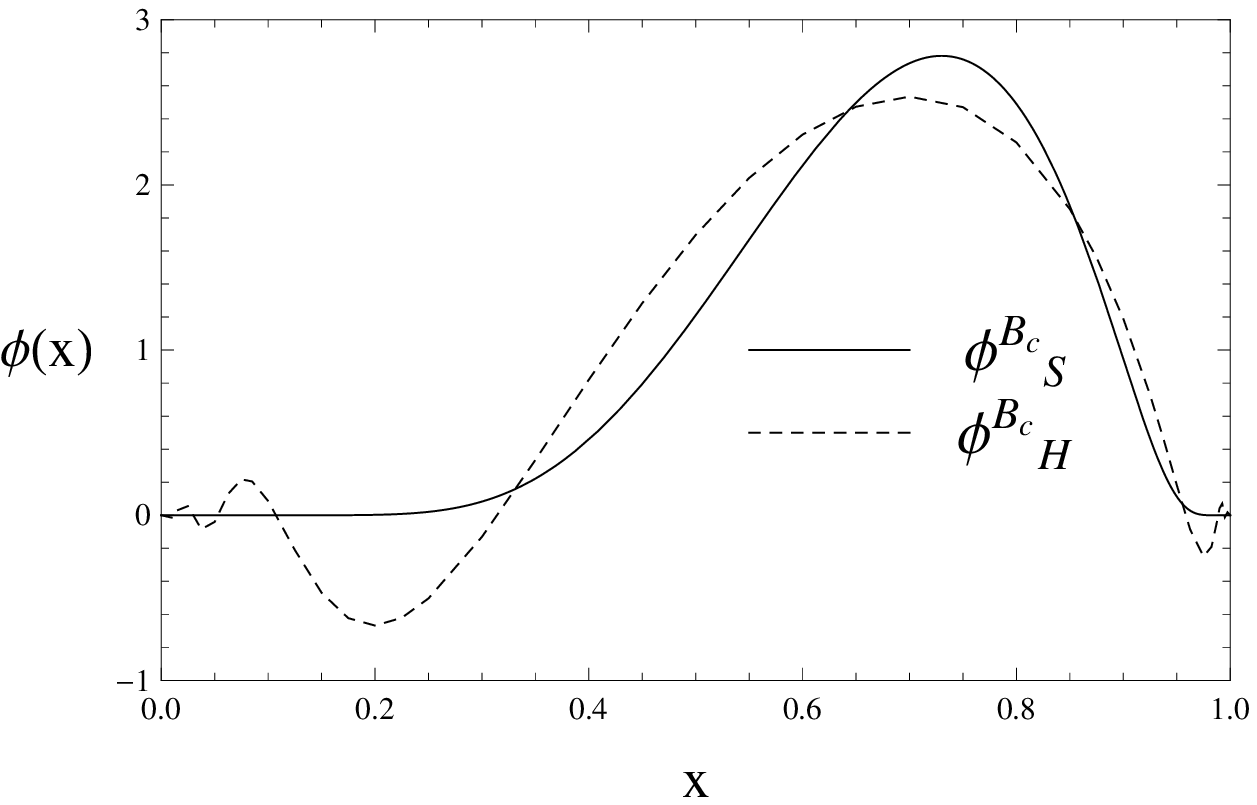}
 \caption{Distribution amplitudes of $B_c$ meson. The solid and dotted lines correspond to the SW and HW models,
 respectively.}
  \label{fig:Bc}
 \end{figure}

For the other light and heavy mesons, we apply the values of current quark mass as $m_u=2.5$ MeV, $m_d=5.0$ MeV,
$m_s=100$ MeV, $m_c=1.29$ GeV, and $m_b=4.19$ GeV \cite{data}. As for the dilaton scale parameter $\kappa$, we try to
let it change with the decay constant of the different mesons. Here we use the following values:
$f_{K^-}=156.1$ MeV, $f_{D^+}=206.7$ MeV, $f_{D_s^+}=257.5$ MeV, $f_{B_d}=193$ MeV,
$f_{B_s}=253$ MeV \cite{data}, $f_{\eta_c}=335$ MeV \cite{PRL}, $f_{B_c}=489$ MeV, and $f_{\eta_b}=801$ MeV \cite{TWQCD}.
From the above input values, the parameters $\kappa$ and $l$ can be fixed as shown in Table \ref{kappa_l}.
\begin{table}[ht!]
\caption{\label{kappa_l} Parameters $\kappa$ and $l$ for the various mesons }
\begin{tabular}{|c|c|c|c|c|c|c|c|c|c|}\hline
 &$\pi$ & $K$ & $D$ & $D_s$ & $B$  & $B_s$ & $\eta_c$ & $\eta_b$ & $B_c$  \\ \hline
 $\kappa$ (GeV) & $0.426$ & $0.503$ & $0.909$ & $1.03$ & $1.49$ & $1.67$ &
   $1.18$ & $2.99$ & $2.08$ \\
 $l$ (fm) & $0.892$ & $0.745$ & $0.337$ & $0.306$ & $0.143$ & $0.139$ &
   $0.222$ & $0.0751$ & $0.112$ \\  \hline
\end{tabular}
\end{table}
We find that, except for the $\pi$ and $K$ mesons, the parameter relation $\kappa=1/l J_1(\beta_{0,1})$ is no longer
satisfied.
For the light meson, the last subsection shows that $\langle r^2\rangle^{\textrm{S}}\propto \kappa^{-2}$ and
$\langle r^2\rangle^{\textrm{H}}\propto l^2$. We compare the experimental data:
$ \langle r^2_\pi\rangle/\langle r^2_{K^+}\rangle\simeq1.44$ \cite{data} with the ratio:
 \be
 \frac{\kappa^2_K}{\kappa^2_\pi}\simeq1.40,\qquad
 \frac{l^2_\pi}{l^2_K}\simeq1.43.
 \en
In addition, for the heavy-light meson, we compare the square root of
heavy quark ratio $ \sqrt{m_c/m_b}\simeq 0.555$ \cite{data} with the ratios
 \be
 \frac{\kappa_{D}}{\kappa_{B}}\simeq0.611,\qquad
 \frac{\kappa_{D_s}}{\kappa_{B_s}}\simeq0.618.
 \en
Thus, we can conclude that these parameters are consistent with the
data and satisfy the result of HQET, Eq. (\ref{HQET}). Next, we use
these parameters to evaluate the distribution amplitude of the mesons.
The normalized distribution amplitudes of the light mesons $(\pi,K)$
for both models are plotted in Fig. \ref{fig:light}.
As shown in Eq. (\ref{comparephi}) for the light meson, the
normalized distribution amplitudes for the SW and HW models are
nearly the same. As for the other mesons, the normalized
distribution amplitudes of the heavy-light mesons $D$, $D_s$, $B$,
and $B_s$ for both models are plotted in Figs. \ref{fig:D} ,
\ref{fig:Ds}, \ref{fig:B}, and \ref{fig:Bs}, respectively.
The normalized distribution amplitudes of the heavy quarkonium states $\eta_c$, $\eta_b$,
and $B_c$ for both models are plotted in Figs. \ref{fig:etac} , \ref{fig:etab}, and \ref{fig:Bc}, respectively.
These figures show that the normalized distribution amplitudes of the heavy-light and heavy quarkonium mesons
for the HW and SW models are discriminating. Finally, we evaluate the first four $\xi$-moments of these distribution
functions, $\langle\xi^N\rangle^{\textrm{S,H}}$, as defined in Eq. (\ref{ximoment}), and compare the results
with the other theoretical calculations in Table \ref{xi_moment}.
 \begin{table}[ht!]
 \caption{\label{xi_moment} First moments of the distribution function $\langle \xi^N \rangle$ for the various mesons. }
 {\footnotesize
 \begin{tabular}{|c|c|c|c|c|c|c|c|c|c|}\hline
 &$\pi$ & $K$ & $D$ & $D_s$ & $B$  & $B_s$ & $\eta_c$ & $\eta_b$ & $B_c$  \\ \hline
 $\langle \xi^1 \rangle^{\textrm{S}}$ & $0$ & $0.0273$ & $0.385$ & $0.332$ & $0.645$ & $0.597$ &
   $0$ & $0$ & $0.360$ \\
 $\langle \xi^1 \rangle^{\textrm{H}}$ & $0$ & $0.0232$ & $0.402$ & $0.333$ & $0.851$ & $0.726$ &
   $0$ & $0$ & $0.397$ \\
 $\langle \xi^1 \rangle$ \cite{BTmassive} & $0$ & $0.04\pm 0.02$ & $0.71$ &  & $0.96$ &  &
   $0$ & $0$ &  \\
 $\langle \xi^1 \rangle$ \cite{lattice1} & $0$ & $0.029\pm 0.002$ &  &  &  &  &
    &  &  \\
 $\langle \xi^1 \rangle$ \cite{lattice2} & $0$ & $0.0272\pm 0.0005$ &  &  &  &  &
    &  &  \\
 $\langle \xi^1 \rangle^g$ \cite{hwangepc,hwangprd} &  &  & $0.288$ & $0.213$ & $0.617$ & $0.549$ &
   $0$ & $0$ & $0.536$\\ \hline
 $\langle \xi^2 \rangle^{\textrm{S}}$ & $0.250$ & $0.239$ & $0.271$ & $0.245$ & $0.469$ & $0.421$ &
   $0.0957$ & $0.0726$ & $0.202$ \\
 $\langle \xi^2 \rangle^{\textrm{H}}$ & $0.250$ & $0.240$ & $0.245$ & $0$.222 & $0.479$ & $0.395$ &
   $0.0924$ & $0.0575$ & $0.170$ \\
 $\langle \xi^2 \rangle$ \cite{BTmassive} & $0.25$ & $0.235\pm 0.005$ & $0.54$ &  & $0.91$ &  &
   $0.02$ & $0.002$ &  \\
 $\langle \xi^2 \rangle$ \cite{lattice1} & $0.28\pm 0.03$ & $0.27\pm 0.02$ &  &  &  &  &
    &  &  \\
 $\langle \xi^2 \rangle$ \cite{lattice2} & $0.269\pm 0.039$ & $0.260\pm 0.006$ &  &  &  &  &
    &  &  \\
 $\langle \xi^2 \rangle^g$ \cite{hwangepc,hwangprd} &  &  & $0.210$ & $0.183$ & $0.425$ & $0.359$ &
  $0.117$  & $0.0643$ & $0.227$\\
 $\langle \xi^2 \rangle$ \cite{Braguta} &  &  &  &  &  &  &
   $0.070\pm 0.007$ &  & \\
 $\langle \xi^4 \rangle$ \cite{Bell} &  &  &  &  &  &  &
   $0.067$ &  & \\ \hline
 $\langle \xi^3 \rangle^{\textrm{S}}$ & $0$ & $0.0180$ & $0.182$ & $0.154$ & $0.362$ & $0.316$ &
   $0$ & $0$ & $0.117$ \\
 $\langle \xi^3 \rangle^{\textrm{H}}$ & $0$ & $0.0156$ & $0.180$ & $0.155$ & $0.385$ & $0.328$ &
   $0$ & $0$ & $0.130$ \\
 $\langle \xi^3 \rangle^g$ \cite{hwangepc,hwangprd} &  &  & $0.125$ & $0.0890$ & $0.312$ & $0.254$ &
   $0$ & $0$ & $0.133$\\ \hline
 $\langle \xi^4 \rangle^{\textrm{S}}$ & $0.125$ & $0.115$ & $0.142$ & $0.121$ & $0.290$ & $0.249$ &
   $0.0216$ & $0.0129$ & $0.0754$ \\
 $\langle \xi^4 \rangle^{\textrm{H}}$ & $0.125$ & $0.116$ & $0.126$ & $0.109$ & $0.283$ & $0.232$ &
   $0.0121$ & $-0.00554$ & $0.0690$ \\
 $\langle \xi^4 \rangle^g$ \cite{hwangepc,hwangprd} &  &  & $0.0960$ & $0.0738$ & $0.240$  & $0.189$  &
  $0.0307$  & $0.0103$ & $0.108$\\
 $\langle \xi^4 \rangle$ \cite{Braguta} &  &  &  &  &  &  &
   $0.012\pm 0.002$ &  & \\
 $\langle \xi^4 \rangle$ \cite{Bell} &  &  &  &  &  &  &
   $0.011$ &  & \\ \hline
 \end{tabular}
 }
 \end{table}
\section{Conclusions}
We have compared two types of wave functions for pseudoscalar mesons in the light-front framework,
obtained by the AdS/CFT correspondence within the hard-wall and soft-wall holographic models.
In the case of massless constituents, we find that the asymptotic behaviors of $Q^2 F_\pi(Q^2)$,
the distribution amplitudes, and the decay constants for both models are the same if a parametric
relation, $\kappa=1/l J_1(\beta_{0,1})$, is set. Furthermore, in terms of the normalized wave functions
of the SW model as a complete set, the ground state of the SW wave function dominates within
that of the HW one. On the other hand, by introducing a quark mass dependence, the differences of the
distribution amplitudes between the two models are obvious, and the above parametric relation is no longer
satisfied if the decay constants of the various mesons are regarded as inputs. In addition, for the SW
model, the dependences of the decay constants of meson on the dilaton scale parameter differ:
$f_{qq}\sim \kappa +\mathcal{O}(m_q^2/\kappa)$, $f_{QQ}\sim \kappa^{3/2}/m_Q^{1/2}$, and $f_{Qq}\sim
\kappa^3/m_Q^2$. The last one, if $\kappa\sim m_Q^{1/2}$, is consistent with HQET: $f_{Qq}\sim 1/m_Q^{1/2}$.
Thus, we fit the values of $\kappa$ and $l$ with the decay constants of the distinct mesons and find that
the ratios of parameters are consistent with the prediction of HQET and with the ratios of the mean square
radius for the light mesons. Finally, we plot the distribution amplitudes of mesons for the two models
and compare the first four $\xi$-moments of our estimations with those of the other theoretical
calculations.

{\bf Acknowledgements}\\
This work was supported in part by the National Science
Council of the Republic of China under Grant No
NSC-99-2112-M-017-002-MY3.

\appendix
\section{EM form factor in large $Q$ limit }
\begin{figure}
 \includegraphics*[width=4in]{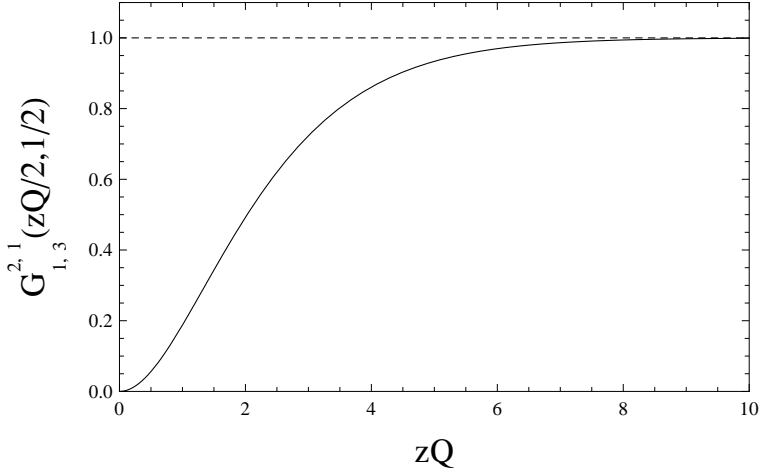}
 \caption{Meijer G function $G^{2,1}_{1,3}(z Q/2,1/2)$. The dotted line equals 1.}
  \label{fig:G}
 \end{figure}

From Eq. (\ref{Fz}), we have the EM form factor of pion as:
 \be
 F_\pi(Q^2)=R^3\int^u_0\frac{dz}{z^3}e^{\varphi(z)}\Phi(z)V(Q^2,z)\Phi(z),\label{A1}
 \en
where $u=l(\infty)$, $\varphi(z)=0(-\kappa^2 z^2)$, and $V(Q^2,z)=V_\textrm{H}(Q^2,z)
(V_\textrm{S}(Q^2,z))$ for the HW (SW) model. Recalling Eq. (\ref{LQ}), that is, $V_\textrm{S}(Q^2,z)\rightarrow
V_\textrm{H}(Q^2,z)=z Q K_1(z Q)$ in the large $Q$ limit, we rewrite Eq. (\ref{A1}) as:
 \be
 F_\pi(Q^2)=\frac{R^3}{Q^2}\int^{Q u}_0\frac{e^{\varphi(z)}\Phi(z)^2}{z^4}(z Q)^2K_1(z Q)d(z Q),\label{A2}
 \en
The integral Eq. (\ref{A2}) can be taken the integration by parts, and the result is:
 \be
 F_\pi(Q^2)=\frac{R^3}{Q^2}\bigg[\Psi(z) G^{2,1}_{1,3}\bigg(\frac{z Q}{2},
 \frac{1}{2}\bigg|\begin{array}{lll}
 1&&\\1,&2,&0\end{array}\bigg)\bigg|^{Q u}_0
 -\int^{Q u}_0\Psi'(z) G^{2,1}_{1,3}\bigg(\frac{z
 Q}{2},\frac{1}{2}\bigg|
 \begin{array}{lll}1&&\\1,&2,&0\end{array}\bigg)d(z Q)
 \bigg],\non \\ \label{A3}
 \en
where $\Psi(z)\equiv e^{\varphi(z)}\Phi(z)^2/z^4$, $\Psi'(z)=d\Psi(z)/d(z Q)$, and $G^{m,n}_{p,q}(z,r)$ is the generalized form
of Meijer G function. We plot the curve of this Meijer G function
in Fig. \ref{fig:G} and find it approaches $1$ and $0$ in the large $Q$ limit and $Q\rightarrow0$, respectively. Thus, the first
term of Eq. (\ref{A3}) vanishes because $\Phi(z)\sim z^2$ for both two models, and the second term approximately equals a constant when the large $Q$ limit is taken:
 \be
 F_\pi(Q^2)|_{\textrm{large}~Q}\simeq\frac{R^3}{Q^2}\times
 (-)\int^{Q u}_0\Psi'(z) d(zQ)= -\frac{R^3}{Q^2}\frac{e^{\varphi(z)}\Phi(z)^2}{z^4}\Bigg|^{Q u}_0.
 \en
Substituting the relevant functions and numbers, we find:
 \be
 Q^2F^\textrm{H}_\pi(Q^2)|_{Q^2\rightarrow\infty}=4\left(\frac{1}{l J_1(\beta_{0,1})}\right)^2,\quad\quad
 Q^2F^\textrm{S}_\pi(Q^2)|_{Q^2\rightarrow\infty}=4\kappa^2.
 \en
\section{Correlation between $c_n$ and $\kappa$ }
The solutions of the effective
Schrodinger equations, Eqs. (\ref{phieq}) and (\ref{phieqsoft}), for $L=0$ are:
 \be
 \tilde{\Phi}^n_{\textrm{H}}(\zeta)=\frac{\sqrt{2\zeta}}{l J_1(\beta_{0,n+1})}
 J_0\left(\frac{\zeta\beta_{0,n+1}}{l}\right),
 \en
and
 \be
 \tilde{\Phi}^n_{\textrm{S}}(\zeta)=\sqrt{2\zeta}\kappa e^{-\kappa^2\zeta^2/2}L^0_n(\kappa^2\zeta^2),
 \en
where $\int d\zeta |\tilde{\Phi}_{\textrm{H,S}}(\zeta)|^2=1$ is satisfied. If Eq. (\ref{kappaLambda}) is
substituted to Eq. (\ref{expand}), we have:
 \be
 c_n=\int^{1/\kappa J_1(\beta_{0,1})}_0 d\zeta \sqrt{2\zeta}\kappa
 J_0(\zeta\kappa\beta_{0,1}J_1(\beta_{0,1}))\times \sqrt{2\zeta}\kappa e^{-\kappa^2\zeta^2/2}L^0_n
 (\kappa^2\zeta^2). \label{dot}
 \en
We may introduce a variable $\lambda=\kappa^2\zeta^2$ , then $\tilde{\Phi}_{\textrm{H,S}}(\zeta)$ can be rewritten as
 \be
 \tilde{\Phi}'^n_{\textrm{H}}(\lambda)&=&J_0(\sqrt{\lambda}\beta_{0,n+1}J_1(\beta_{0,n+1})),\non \\
 \tilde{\Phi}'^n_{\textrm{S}}(\lambda)&=&e^{-\lambda/2}L^0_n(\lambda),
 \en
which satisfy the normalization condition $\int d\lambda |\tilde{\Phi}'^n(\lambda)|^2=1$. It is well known
that the associated Laguerre polynomials $L^0_n(\lambda)$ have an orthogonality:
 \be
 \int^\infty_0 e^{-\lambda}L^0_n(\lambda)L^0_m(\lambda)d\lambda=\delta_{m,n}.
 \en
Thus, $\{\tilde{\Phi}'^n_{\textrm{S}}(\lambda)\}$ is a complete set and Eq. (\ref{dot}) can be rewritten as:
 \be
 c_n=\int^{1/J_1(\beta_{0,1})^2}_0 d\lambda
 J_0(\sqrt{\lambda}\beta_{0,1}J_1(\beta_{0,1}))\times e^{-\lambda/2}L^0_n
 (\lambda). \label{dotr}
 \en
It is obvious that $c_n$ is only dependent of $\beta$ and is independent of the parameters $\kappa$ and $l$.

\section{Scalings of $f_{QQ}$ and $f_{qq}$}
In considering the heavy quarkonium, the LF wave function for the SW model is:
 \be
 \psi_{\textrm{S}}(x,\textbf{k}_\bot,m_i)=N_{QQ}\frac{4\pi}{\kappa\sqrt{x(1-x)}}\textrm{exp}\left[-\frac{1}{2\kappa^2}
 \left(\frac{k^2}{x (1-x)}+\frac{m_Q^2}{x (1-x)}\right)\right],
 \en
where $N_{QQ}$ is the normalization constant and
 \be
 N_{QQ}=\left[\int^1_0 dx~ \textrm{exp}
 \left(\frac{-m_Q^2}{x (1-x)\kappa^2}\right)\right]^{-1/2}.
 \en
If we change the variable $\sin\theta=1-2 x$, the normalization constant is replaced as:
 \be
 N_{QQ}=\left[\int^{\pi/2}_{-\pi/2} d\theta~ \frac{\cos\theta}{2}\textrm{exp}
 \left(\frac{-4 m_Q^2}{\cos^2\theta\kappa^2}\right)\right]^{-1/2}=\left[\frac{\sqrt{\pi}}{2}
 G^{2,0}_{1,2}\Bigg(\frac{4m_Q^2}{\kappa^2}\Bigg|
 \begin{array}{lll}&\frac{3}{2}&\\0,&1&\end{array}\Bigg)\right]^{-1/2}.
 \en
The distribution amplitude and the decay constant can be evaluated as:
 \be
 \phi_{\textrm{S}}(x,m_Q)=\left[2\sqrt{\pi}
 G^{2,0}_{1,2}\Bigg(\frac{4m_Q^2}{\kappa^2}\Bigg|
 \begin{array}{lll}&\frac{3}{2}&\\0,&1&\end{array}\Bigg)\right]^{-1/2} \frac{\sqrt{x (1-x)} \kappa}
 {\pi}e^{-\frac{m_Q^2}{2x (1-x)\kappa^2}},
 \en
and
 \be
 f_{QQ}=\left[
 G^{2,0}_{1,2}\Bigg(4r^2\Bigg|
 \begin{array}{lll}&\frac{3}{2}&\\0,&1&\end{array}\Bigg)\right]^{-1/2}\frac{\sqrt{3}}{4 \pi^{3/4}}
 \left[2\sqrt{2} r e^{-2 r^2}+\sqrt{\pi}(1-4 r^2) \textrm{erfc}(\sqrt{2} r)\right],\label{fQQ}
 \en
where $r=m_Q/\kappa$ and $\textrm{erfc}(a)$ is the complementary error function.
Performing an expansion in powers of $1/r$ for Eq. (\ref{fQQ}), we obtain:
 \be
 f_{QQ}= \frac{\sqrt{3}}{2\pi^{3/4}} \frac{\kappa}{r^{1/2}}+\mathcal{O}\left(\frac{1}{r^{5/2}}\right).
 \en
Taking the heavy quark limit, we have the scaling: $f_{QQ}\sim \kappa^{3/2}/m_Q^{1/2}$.
For the light quarkonium, we perform an expansion in powers of $r$ for Eq. (\ref{fQQ}) and obtain:
 \be
  f_{qq}= \frac{\sqrt{6}\kappa}{8}-\frac{\kappa}{4}\sqrt{\frac{3}{2}}\left(3+2\gamma+2 \ln[2 r]+\Gamma'
  \left[\frac{1}{2}\right]\right) r^2+\mathcal{O}\left(r^3\right),
 \en
where $\gamma$ is the Euler's constant and $\Gamma'[a]$ is the digamma function.
\section{Scaling of $f_{Qq}$}
In considering the heavy-light meson, the LF wave function for the SW model is:
 \be
 \psi_{\textrm{S}}(x,\textbf{k}_\bot,m_i)=N_{Qq}\frac{4\pi}{\kappa\sqrt{x(1-x)}}\textrm{exp}\left[-\frac{1}{2\kappa^2}
 \left(\frac{k^2}{x (1-x)}+\frac{m_Q^2}{1-x}+\frac{m_q^2}{x}\right)\right],
 \en
where $N_{Qq}$ is the normalization constant and
 \be
 N_{Qq}=\left\{\int^1_0 dx ~\textrm{exp}\left[-\frac{1}{\kappa^2}
 \left(\frac{m_Q^2}{1-x}+\frac{m_q^2}{x}\right)\right]\right\}^{-1/2}.
 \en
Taking the limit $m_Q\gg m_q$ and ignoring the light quark mass, we obtain:
 \be
 N_{Qq}\simeq\left\{e^{-\frac{m_Q^2}{\kappa^2}}-\frac{m_Q^2}{\kappa^2}\Gamma[0,\frac{m_Q^2}{\kappa^2}]\right\}^{-1/2},
 \en
where $\Gamma[a,z]$ is the incomplete gamma function. Thus, the distribution amplitude and the decay constant can be evaluated as:
 \be
 \phi_{\textrm{S}}(x,m_Q)=\frac{\sqrt{x (1-x)} \kappa}{2\pi\left\{e^{-\frac{m_Q^2}{\kappa^2}}-\frac{m_Q^2}{\kappa^2}\Gamma[0,\frac{m_Q^2}
 {\kappa^2}]\right\}^{1/2}}e^{-\frac{m_Q^2}{2(1-x)\kappa^2}},
 \en
and
 \be
 f_{Qq}\simeq \frac{\sqrt{3}\kappa}{4\sqrt{2\pi}}\frac{\sqrt{2} r(r^2+1)e^{-\frac{r^2}{2}}-\sqrt{\pi} (r^4+2 r^2-1)
 \textrm{erfc}\left(\frac{r}{\sqrt{2}}\right)}{\left\{e^{-r^2}-r^2
 \Gamma[0,r^2]\right\}^{1/2}},\label{fQq}
 \en
We perform an expansion
in powers of $1/r$ for Eq. (\ref{fQq}) and obtain:
 \be
 f_{Qq}\simeq 2\sqrt{\frac{3}{\pi}} \frac{\kappa}{r^2}+\mathcal{O}\left(\frac{1}{r^3}\right).
 \en
Taking the heavy quark limit, we have the scaling: $f_{Qq}\sim \kappa^3/m_Q^2$.

\end{document}